\documentclass[11pt,twoside]{article}
\usepackage{asp2004}
\usepackage{epsf}
\usepackage{graphics}
\usepackage{graphicx}
\usepackage{lscape}
\usepackage{multirow} 
\begin{document}

\title{ New release of the ELODIE library: Version 3.1}
\author{Ph. Prugniel$^{(1,2)}$, C. Soubiran$^{(3)}$, M. Koleva$^{(1,4)}$ \& D. Le Borgne$^{(5)}$}
\affil{(1) Universit\'e Lyon~1, CRAL-Observatoire de Lyon, France,\\
(2) GEPI-Observatoire de Paris, France,\\
(3) LAB-Observatoire de Bordeaux, France, \\
(4) Department of Astronomy, St. Kl. Ohridski University of Sofia, BG-1164 Sofia, Bulgaria,\\
(5) Service d'Astrophysique, CEA-Saclay, France}

\begin{abstract}
 We present ELODIE.3.1, an updated release of the library  published in 
 Prugniel \& Soubiran (2001, 2004). 
 The library includes 1962 spectra of  1388 stars obtained with the ELODIE
 spectrograph at the Observatoire de  Haute-Provence 193cm telescope in
 the wavelength range 390 to 680 nm. It  provides a wide coverage of
 atmospheric parameters : T$_{eff}$ from 3100 K  to 50000 K, $log g$ from -0.25
 to 4.9 and [Fe/H] from -3 to +1. The  library is given at two
 resolutions: R$\approx$42000, with the flux normalized  to the
 pseudo-continuum, FWHM=0.55\AA (R$\approx$10000) calibrated in physical flux 
 (reduced  above earth atmosphere) with a broad-band photometric precision of 
 2.5\%  and narrow-band precision of 0.5\%. 

 In this new release the data-reduction (flux calibration, reconnection
 of the echelle orders) has been improved, and in particular the blue region,
 between 390 and 400 nm has been added.

 The FITS files for each spectra, and the measured atmospheric parameters
 are publicly available.
 See the ELODIE.3.1 page for more details:\\ 
 
 \noindent
 http://www.obs.u-bordeaux1.fr/m2a/soubiran/elodie\_library.html

 \noindent
 Email: prugniel@obs.univ-lyon1.fr

\end{abstract}
\thispagestyle{plain}

\section{Introduction}

\begin{figure*}[ht]
\centering
\includegraphics[width=11cm]{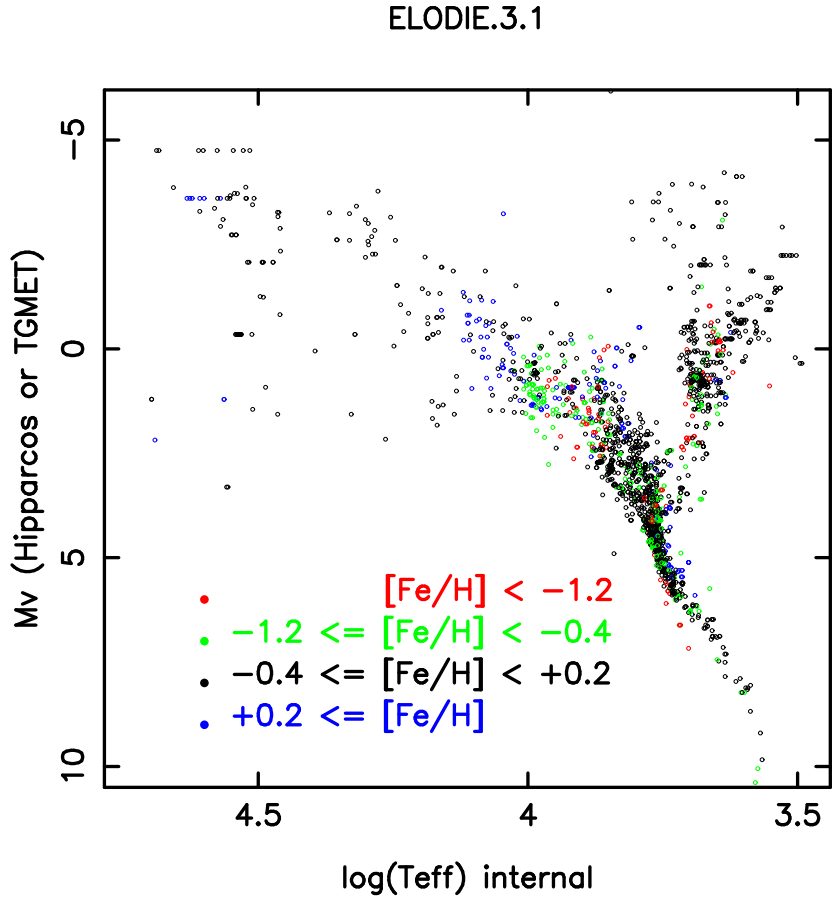}
\caption{Distribution of the ELODIE.3.1 library
in the HR diagram, with three metallicity bins differentiated with symbols 
of different colors
}
\label{TeffMv}
\end{figure*}

The ELODIE library is based on spectra retrieved from the Observatoire
de Haute-Provence archive \footnote{http://atlas.obs-hp.fr/elodie}
(Moultaka et al. 2004). The spectra were obtained with the ELODIE
spectrograph attached to the 1.93m telescope, they have a spectral
resolution of about R=42000 and consists of 67 echelle orders covering
in total the wavelength range 389.2 to 680 nm without any gap.

Some of the spectra were acquired for the purpose of building a stellar 
library, and others were selected from the archive to complete
the coverage in effective temperature (T$_{eff}$), surface gravity
($log g$) and metallicity ([Fe/H]) and to achieve the flux calibration
(by providing enough external comparisons and repeated observations
of the same stars). 

An early library, not calibrated in flux, was published in Soubiran et al.
(1998). The first flux-calibrated release of the library was published
by Prugniel \& Soubiran (2001), it contained 906 spectra.
The last previous release, ELODIE.3 was announced in Prugniel \& Soubiran
(2004) and represented a major progress in the coverage of the space of
atmospheric parameters (T$_{eff}$, $log g$ and [Fe/H]).

The present version, ELODIE.3.1 is based on the same collection of
spectra than ELODIE.3.
Fig. 1, 2 and 3 show the distribution of the 1388 stars of the library in
the HR diagram, T$_{eff}$ - [Fe/H] and  T$_{eff}$ - $log g$ planes. 
The main interest of this new version is the extended wavelength coverage,
now including the H \& K lines, and limited by the blue limit of
the spectrograph. Some important improvements are also summarized in the
next section.

The library has been used for population synthesis using the Pegase.HR program
(Le Borgne et al. 2004), and these models were used to study the stellar
populations of galaxies and star clusters. The advantage of these models is
the high spectral resolution provided by the library which allows to
constrain simultaneously the internal kinematics and the parameters
of the stellar population (for instance age and metallicity). Thanks
to this library, the analysis is not anymore limited to the low resolution
of the spectrophotometric indices. 
Specific methods for the simultaneous analysis of the kinematics and of the 
stellar population were developed in Ocvirk et al. (2006) and Chilingarian 
(2007). It was found that these approaches of full spectrum fitting
are increasing the precision on age and metallicity by a factor 3
(a factor 10 on observing time to get the same precision) (Koleva et al. 2006).

The comparison between models based on the current release of the library
and other models are presented in Koleva et al (2007a) and validation
using Galactic clusters in Koleva et al. (2007b).

We are now preparing a further version, ELODIE.4, which will contain about
5000 spectra. This will allow a better investigation of the effect of
the $\alpha$-element abundances in stellar populations, possibly by
coupling the library with a basis of theoretical spectra as presented
in Prugniel et al. (2007).

\begin{figure*}[ht]
\centering
\includegraphics[width=11cm]{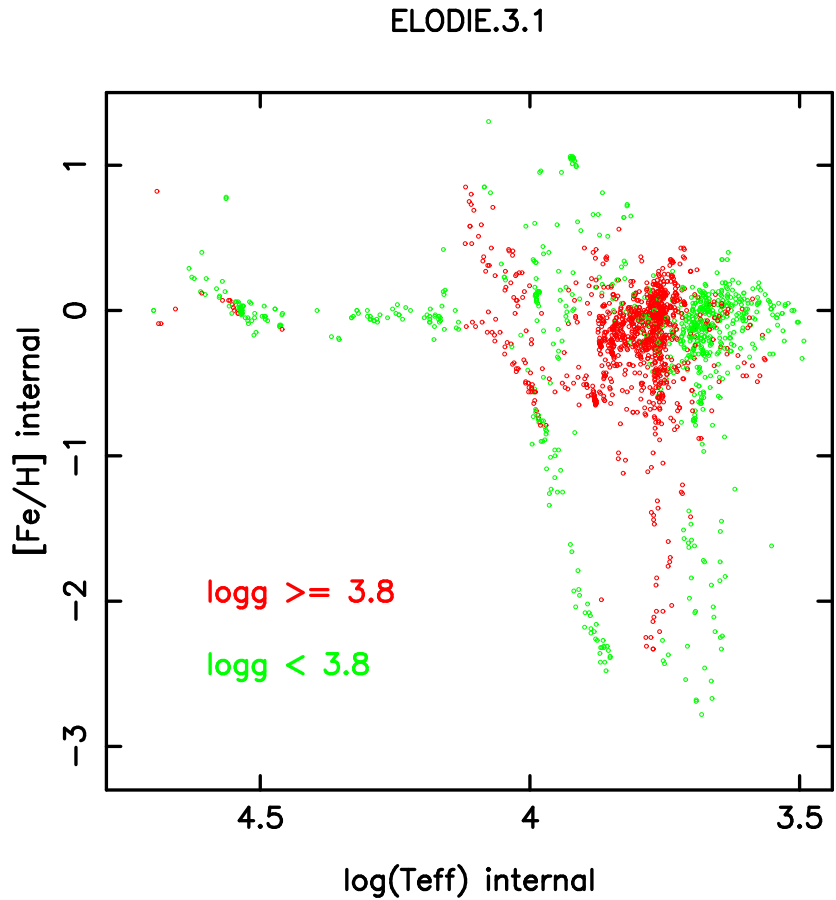}
\caption{Distribution in the T$_{eff}$ - [Fe/H] plane.
Dwarfs with red dots and giants in green.
}
\label{TeffFeH}
\end{figure*}

\begin{figure*}[ht]
\centering
\includegraphics[width=11cm]{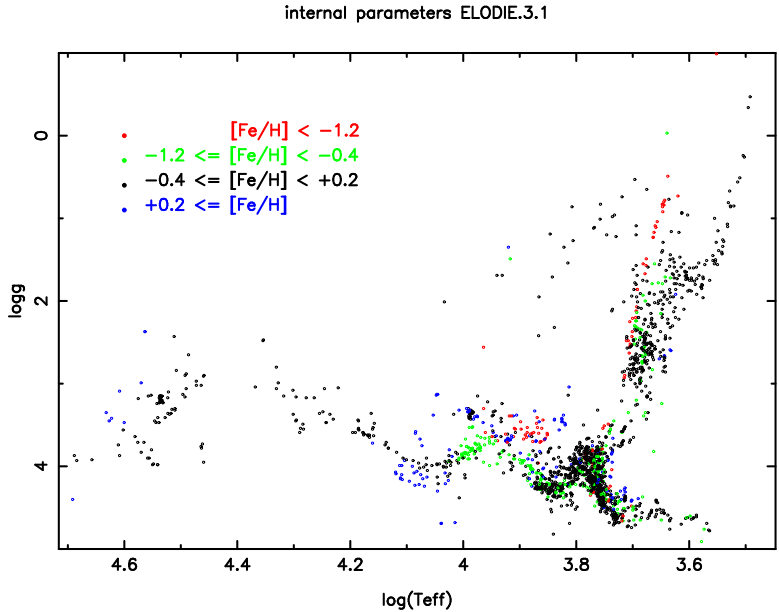}
\caption{Distribution in the T$_{eff}$ - $log g$ plane.
The color of the symbols distinguish the different metallicity classes.
}
\label{Tefflogg}
\end{figure*}

\begin{figure*}[ht]
\centering
\includegraphics[width=11cm]{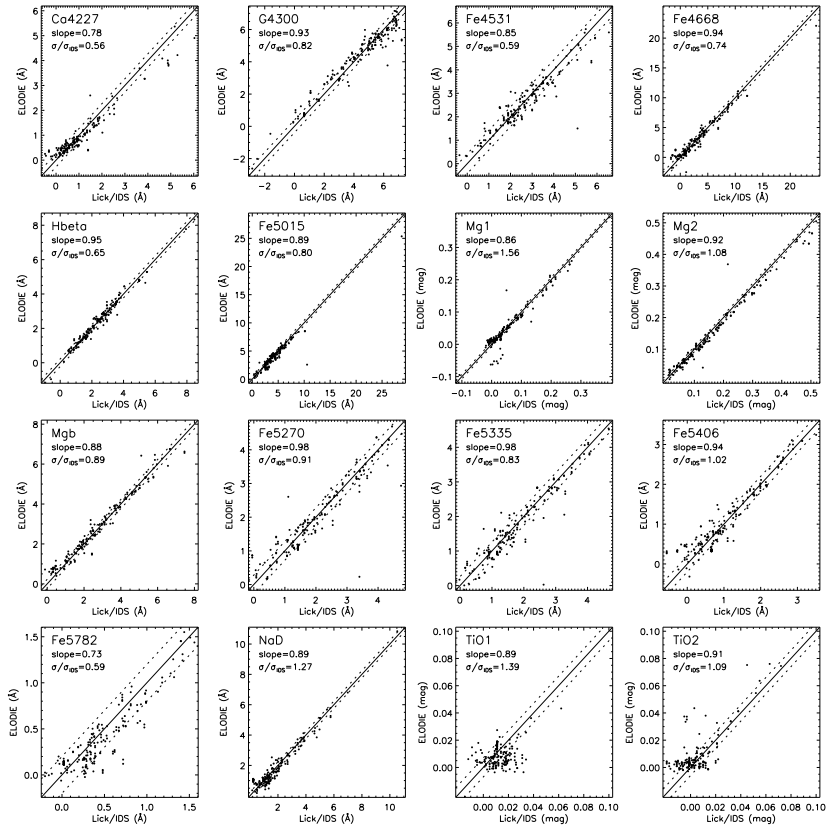}
\caption{ 
Comparison between Lick indices measured on ELODIE library and the 
Worthey et al. (1994) values.
}
\label{lickworthey}
\end{figure*}

\begin{figure*}[ht]
\centering
\includegraphics[width=11cm]{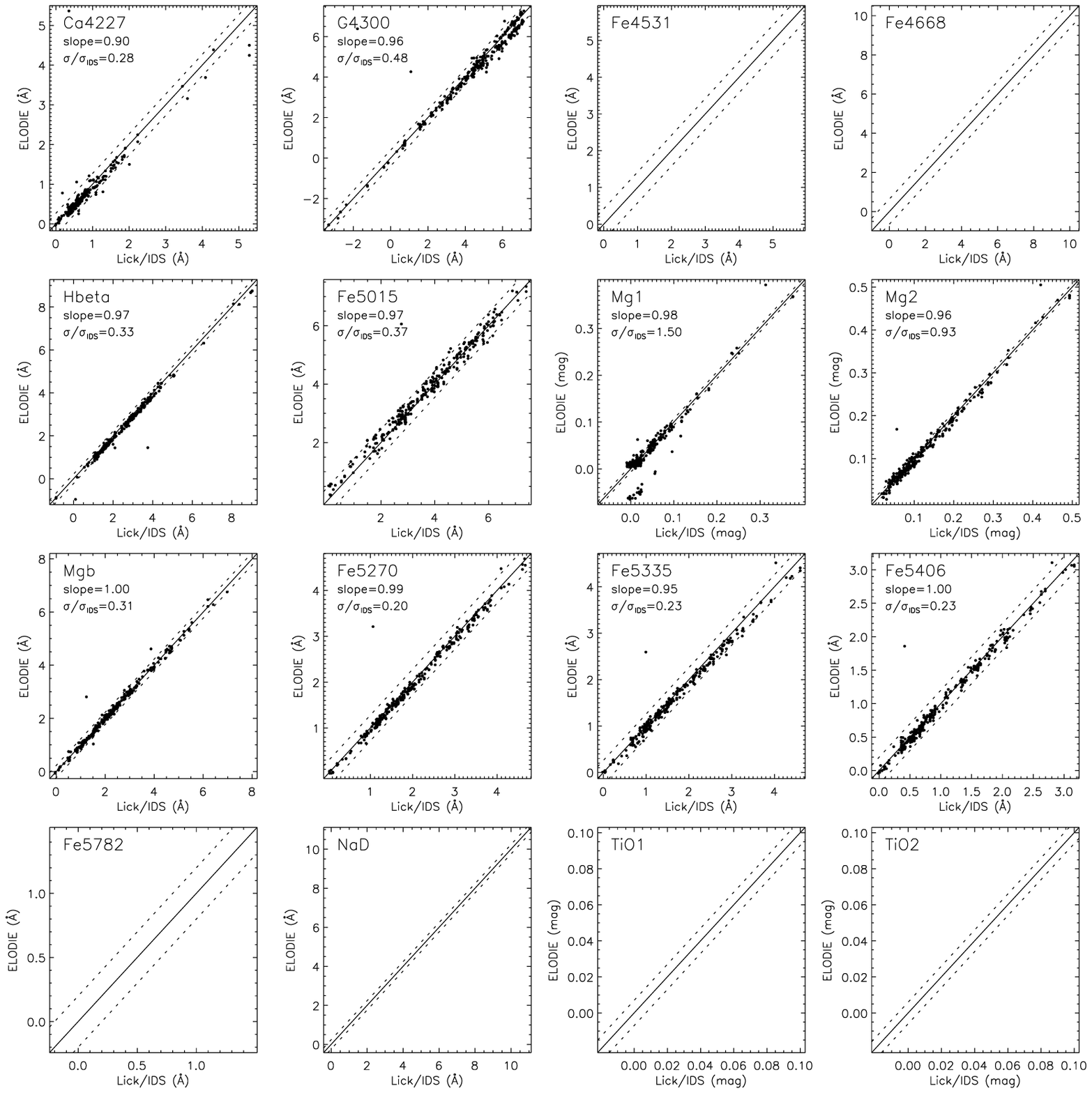}
\caption{ 
Comparison between Lick indices measured on ELODIE library and the Jones 
library values (Worthey \& Ottaviani, 1997)
}
\label{lickjones}
\end{figure*}

\section{ Reduction procedure}
Since the first version of the library, the reduction procedure described 
in Prugniel \& Soubiran (2001) has been improved in several aspects but the 
general philosophy remains the same. 
The first steps of the data processing up to the extraction of the orders
are made with the standard ELODIE pipeline (Baranne et al., 1996).
The reconnection of the orders, and the flux calibration are explained
in  Prugniel \& Soubiran (2001).

The specific points modified in the current release are:

\begin{description}

\item[Extension of the wavelength range.]
In the previous releases the first 4 orders in the blue were not processed
because of the difficulty of their reconnection and calibration due
to the low signal-to-noise ratio (S/N). The sensitivity of the detector is
indeed dramatically dropping near the blue limit.
We improved our procedure and we could include the blue region, extending
to the H \& K lines.

\item[Subtraction of the diffuse light.]
When working on the blue region of the spectra, we discovered that the diffuse 
light on the standard pipeline was often under-subtracted. The consequence
was that some sharp lines in the blue were not deep enough.
The correction that we applied was validated by comparing with spectra
from other archives or libraries.

\item[Improvement of the atmospheric parameters.]
To use the library for population synthesis, a key factor is to obtain the
atmospheric parameters of the stars with the highest accuracy. In the present
case, the atmospheric parameters are first compiled from the literature, and
an inter-comparison in the library allow to detect inconsistencies and
to determine 'internal' values: Each wavelength point of the spectra is
modeled as polynomials in T$_{eff}$, $log g$ and [Fe/H] fitted to
the compiled parameters and the inversion of this function returns internal
determinations of the parameters (see Prugniel \& Soubiran 2001, Le Borgne 
et al. 2004).\\
On one side we have updated the compiled input parameters, using recent
publications and fixing errors, and on the other side we improved the
polynomial functions used to model the spectra. The present version is
better, but, as seen in Fig.~2 and 3, some artifacts persist in the
temperature range 6500-7500K at low metallicity (some input metallicities 
are underestimated).\\
By comparing the observed spectra with the polynomial model, we get an idea of
the quality of the model and of the consistency of the atmospheric parameters.
The rms residuals from the 'absolute' model (ie. using the input atmospheric
parameters) is 2.0\%. The rms residuals from the 'internal' model (iterated
using the atmospheric parameters returned after an inversion with the 
'absolute' model) is 1.4\%. The decrease of the rms residuals after the 
iteration is due to an improvement of the consistency of the atmospheric
parameters. The final residuals include the effect of the noise and of all
the neglected parameters, like rotation, detailed abundances... The
internal determination of the atmospheric parameters may be biased with
respect to the input values: We checked these biases by averaging them
in some regions of the parameters space, and did not find evidence for
clear bias.

\end{description}

Tables 1 \& 2, and Fig. 4 \& 5 compare the Lick indices measured on the
ELODIE spectra with those of the Lick (Worthey et al. 1994) and Jones
(Worthey \& Ottaviani, 1997) libraries. The comparison with the Lick data 
shows an important spread
due to the lower quality of the Lick data, but the comparison with
Jones library reveals an excellent consistency. The zero-points are
extremely small, and the slope between the two series is close to 1.
This test of the slope is critical since it may unveil some data-reduction
errors in the cosmic ray clipping, the rebinning or the scattered light
subtraction: These three aspects potentially affecting more the 
strong lines, hence high metallicity stars.

\begin{table}
\centering
\begin{tabular}{ l  c c  c}
\hline
\hline
    \multirow{2}*{Index}&   \multicolumn{3}{c}{Worthey}    \\
\cline{2-4}
                        &   offset&    slope& $\sigma$     \\
\hline
    Ca4227              &  0.042  &  0.778  &  0.152       \\
     G4300              &  0.019  &  0.930  &  0.321       \\
    Fe4531              &  0.007  &  0.853  &  0.247       \\
    Fe4668              &  0.023  &  0.935  &  0.470       \\
     Hbeta              &  0.022  &  0.951  &  0.143       \\
    Fe5015              & -0.142  &  0.889  &  0.359       \\
       Mg1              & -0.002  &  0.862  &  0.011       \\
       Mg2              &  0.001  &  0.916  &  0.009       \\
       Mgb              & -0.032  &  0.885  &  0.204       \\
    Fe5270              & -0.036  &  0.980  &  0.253       \\
    Fe5335              & -0.019  &  0.975  &  0.215       \\
    Fe5406              & -0.007  &  0.936  &  0.203       \\
    Fe5782              & -0.012  &  0.733  &  0.119       \\
       NaD              &  0.069  &  0.894  &  0.305       \\
      TiO1              & -0.002  &  0.886  &  0.010       \\
      TiO2              &  0.001  &  0.909  &  0.007       \\
\hline
\end{tabular}
\caption{
Comparison with Lick indices from the Lick library. Col. 1: Designation of the Lick index; Col. 2, 3: Index(Elodie) = offset + slope $\times$ Index(Worthey);
Col. 4: rms dispersion from the linear fit.}
\end{table}

\section{Data access}
The spectra are available as fits files. Three gzipped tarfiles, corresponding 
to 
the R=10000, R=42000 libraries can be downloaded, as well as 
the table of measurements
containing the stellar parameters estimated internally and the measured
Lick indices.  
The description of the files and the specific keywords are described
on the ELODIE.3.1 web page.

Two versions of the library are constructed. The high resolution (R $\approx$
42000) corresponds to the nominal resolution of the spectrograph, and the
low resolution (FWHM = 0.55 \AA, R $\approx$ 10000) has been produced
by a gaussian broadening. The high resolution version is provided
normalized to the pseudo-continuum while the low resolution is in 
physical flux with an absolute calibration (physical units) scaled
using Tycho photometry. Accessing the library using the HyperLeda 
database\footnote{http://leda.univ-lyon1.fr}
allows conversions of the flux normalization, or various convolutions
and rebinnings.

Users of this library are kindly requested to cite in their publications
the original A\&A publication (Prugniel \& Soubiran 2001) as well as the 
present astro-ph announcement. The proper denomination of the current
version is: ELODIE.3.1.

We are encouraging reports on the usages of this library and encountered
difficulties. They will help us to continue to improve the quality.

\begin{table}
\centering
\begin{tabular}{ l  c c  c}
\hline
\hline
    \multirow{2}*{Index}&     \multicolumn{3}{c}{Jones}    \\
\cline{2-4}
                        &   offset&    slope& $\sigma$     \\
\hline
    Ca4227              &  0.025  &  0.902  &  0.075       \\
     G4300              &  0.079  &  0.962  &  0.187       \\
     Hbeta              &  0.003  &  0.967  &  0.072       \\
    Fe5015              &  0.006  &  0.967  &  0.166       \\
       Mg1              & -0.003  &  0.978  &  0.011       \\
       Mg2              &  0.000  &  0.961  &  0.007       \\
       Mgb              &  0.021  &  0.997  &  0.072       \\
    Fe5270              &  0.012  &  0.992  &  0.057       \\
    Fe5335              &  0.001  &  0.952  &  0.061       \\
    Fe5406              &  0.008  &  0.998  &  0.047       \\
\hline
\end{tabular}
\caption{
Comparison with Lick indices from the Jones library. Col. 1: Designation of the Lick index; Col. 2, 3: Index(Elodie) = offset + slope $\times$ Index(Jones);
Col. 4: rms dispersion from the linear fit.
}
\end{table}

\noindent {\bf Acknowledgments.}
We thank Observatoire de Haute-Provence, and in particular Sergio Ilovaisky, 
for maintening a public archive of their observations.
Many improvements of this library were possible thanks to the feed-back
of the users.

\end{document}